\documentclass[prb,twocolumn]{revtex4}% Physical Review B
\usepackage{graphicx}% Include figure files
\usepackage{bm}% bold math
\begin{document}

\title{\bf  Anisotropic spin relaxation in quantum dots}
\author{O. Olendski and T. V. Shahbazyan}
\affiliation{Department of Physics, Jackson State University, Jackson,
  Mississippi 39217, USA}

 \begin{abstract}
We study theoretically phonon-assisted spin relaxation of an electron
confined in an elliptical quantum dot (QD) subjected to a tilted magnetic
field. In the presence of both Rashba and Dresselhaus
spin-orbit terms, the relaxation rate is anisotropic with respect to
the in-plane field orientation. This anisotropy originates from the
interference, at nonzero tilt angle, between the two spin-orbit
terms. We show that, in a narrow range of magnetic field orientations,
the relaxation rate exhibits anomalous sensitivity to variations of
the QD parameters. In this range, the relative change in the relaxation
rate with in-plane field orientation is determined {\em solely} by the
spin-orbit coupling strengths and by the dot geometry. This allows 
simultaneous determination of both Rashba and Dresselhaus coupling
parameters and the dot ellipticity from analysis of the angular
dependence  of the relaxation rate.
\end{abstract}
\maketitle

\section{introduction}
\label{intro}

Spin relaxation in semiconductor quantum dots (QDs) has recently
attracted intense interest because of the possible use of the electron
spin as a qubit.\cite{loss-pra98} Quantization of orbital states in a
QD due to the zero-dimensional (0D) confinement leads to the
suppression of  traditional spin-relaxation mechanisms (e.g.,
D'yakonov-Perel) that are dominant in continuous systems. Indeed,
recent experiments on GaAs QDs in a magnetic field $B$ have revealed
extremely long spin-relaxation times (up to $T_1=170$ ms at $B\simeq 1.75$ T).
\cite{tokura-nature02,hanson-prl03,elzerman-nature04,kroutvar-nature04,hanson-prl05,tokura-prl06,kastner-condmat06}
For moderate and high fields ($B > 1$ T), spin relaxation in QDs is
dominated by phonon-assisted electronic transitions between
Zeeman-split levels due to spin-orbit (SO) coupling.
\cite{khaetskii-prb00,lyanda-prb02,golovach-prl04,bulaev-prb05,ulloa-prb05,falko-prl05,konemann-prl05,stano-prl06}
There are two distinct types of SO couplings, one originating
from bulk inversion asymmetry (Dresselhaus coupling) and the other
from structural inversion asymmetry along the growth direction (Rashba
coupling), that cause the admixture of orbital states with opposite
spins. \cite{winkler-book} In the case of circular QDs in a
perpendicular magnetic field, the Rashba and Dresselhaus terms mix
{\em different} pairs of levels, and can, in principle, be
distinguished if one such pair provides the dominant 
relaxation channel. This is the case, for example, when adjacent
orbital levels, coupled via the Rashba term, are brought into resonance
with changing magnetic field.\cite{bulaev-prb05}
However, in more realistic situations, deformations of the QD shape
strongly alter the electronic spectrum \cite{austing-prb99} and, in
general, the effects of the two SO contributions are not separable.
\cite{valin-rodriguez-prb02}

An important distinction between the Rashba and Dresselhaus terms is
their different symmetry properties. The former, described by the
Hamiltonian
$\hat{H}_R=\alpha_R\bigl(\sigma_x\pi_y-\sigma_y\pi_x\bigr)$,
possesses an in-plane rotational symmetry, while the latter,
$\hat{H}_D=\alpha_D\bigl(\sigma_x\pi_x-\sigma_y\pi_y\bigr),$ does
not.\cite{winkler-book}  Here ${\bm \pi}= -i{\bm \nabla}+e{\bf A}$,
${\bf A}$ being the vector potential, ${\bm \sigma}$ is the Pauli
matrix vector, and $\alpha_R$ ($\alpha_D$) is Rashba (Dresselhaus)
coupling constant (we set $\hbar=1$). This lack of rotational
invariance for $\hat{H}_D$ leads to an in-plane momentum 
{\em azimuthal anisotropy} in the presence of both SO terms that
was recently reported in transport experiments in quantum wells.
\cite{marcus-prl03,ganichev-prl04} In a magnetic field, the anisotropy
arises due to the interference between Rashba and Dresselhaus
terms in the matrix elements \cite{falko-prb92} in the presence of
an in-plane field component.  In QDs, this anisotropy reveals itself as
a modulation of the spin relaxation rate for different orientations of
the in-plane field.
\cite{golovach-prl04,falko-prl05,konemann-prl05,stano-prl06}

\begin{figure}
\centering
\includegraphics[width=0.8\columnwidth]{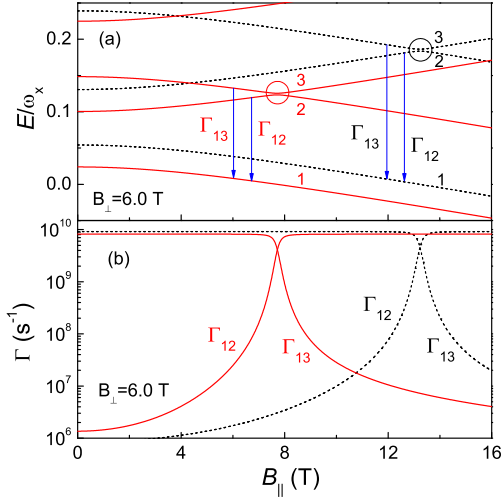}
\caption{\label{fig1}
(Color online) (a) Energy levels and (b) relaxation rates
$\Gamma_{12}$ and $\Gamma_{13}$ for circular (dashed line) and
elliptical (solid line) QDs vs. in-plane field component $B_{||}$.
Arrows in panel (a) indicate relevant transitions, and circles indicate resonance regions.
}
\end{figure}

Here we study the spin relaxation between Zeeman-split levels in
elliptical QDs in a tilted magnetic field. We demonstrate that
the interplay between SO interactions and QD geometry leads to
dramatic changes in the relaxation rate in a certain range of field
orientations for which Rashba and Dresselhaus contributions undergo
{\em destructive} interference. Furthermore, in the vicinity
of level anticrossings (see Fig.~\ref{fig1}), the SO contribution to
the relaxation rate factors out from the phonon
one. This allows simultaneous determination of the parameters for both
SO interactions and QD geometry from the azimuthal anisotropy of
the {\em differential} (with respect to angle) relaxation rate.

The paper is organized as follows. In Section \ref{spin-orbit} we
derive electronic spectrum of elliptical QD in a tilted
magnetic field with both Rashba and Dresselhaus SO terms included. In
Section \ref{phonon} the spin relaxation rate between lowest levels is
evaluated. Numerical results for GaAs dots are presented in Section
\ref{results}.

\section{Spin-Orbit coupling and electron states in elliptical QD in
  tilted magnetic field}
\label{spin-orbit}

We start with the electronic spectrum in an elliptical QD in the presence
of SO interactions subjected to a tilted magnetic field
\begin{equation}
{\bf B}={\bf B}_\perp+{\bf B}_{||}
=B\bigl(\hat{x}\sin\theta\cos\varphi
+\hat{y}\sin\theta\sin\varphi+\hat{z}\cos\theta\bigr),
\end{equation}
where $\theta$ and $\varphi$ are the tilt and azimuthal angles, respectively. The
system is described by the Hamiltonian
$\hat{H}=\hat{H}_0+\hat{H}_{SO}+\hat{H}_Z, $ where $ \hat{H}_0={\bm 
\pi}^2/2m+V_c$ is the Hamiltonian of a 2D electron confined by the parabolic
potential
\begin{equation}
\label{conf}
V_c=m\bigl(\omega_x^2x^2+\omega_y^2y^2\bigr)/2,
\end{equation}
$\omega_x$ and $\omega_y$ being the frequencies of the QD in-plane
potential,
$
\hat{H}_{SO}=\hat{H}_R+\hat{H}_D
$
is the SO term, and
$
\hat{H}_Z=\frac{1}{2}g^\ast\mu_B{\bm \sigma}\cdot{\bf B}
$
is the Zeeman term ($m$, $g^\ast$ and $\mu_B$ stand for the electron effective
mass, $g$ factor and Bohr magneton). For sufficiently strong 2D
confinement, the orbital effect of the in-plane field can be neglected
and $\hat{H}_0$ depends only on the perpendicular field component via
$
{\bf A}=\frac{B_\perp}{2}\bigl(-y,x,0\bigr)
$
(in symmetrical gauge). Using the transformation
\begin{eqnarray}
\label{transf-ellipt}
\label{Pi_x} \pi_x=a_1P_1-a_2P_2, ~
\pi_y=-m\omega_1b_1Q_1+m\omega_2b_2Q_2
\nonumber\\
x=a_1Q_1-a_2Q_2,
~
y=\frac{b_1}{m\omega_1}P_1
-\frac{b_2}{m\omega_2}P_2,
\qquad
\end{eqnarray}
$P_{j}$ and $Q_{j}$ ($j=1,2$) being canonical momenta and coordinates, the
Hamiltonian $\hat{H}_0$ can be brought to the canonical form of  two uncoupled
oscillators with frequencies  \cite{galkin-prb04}
\begin{equation}
\label{frequency}
\omega_{1,2}=\frac{1}{2}\biggl[
\sqrt{\bigl(\omega_x+\omega_y\bigr)^2+\omega_c^2}
\pm\sqrt{\bigl(\omega_x-\omega_y\bigr)^2+\omega_c^2}\biggr],
\end{equation}
where $\omega_c=eB_\perp\bigl/m$ is the cyclotron frequency and
the coefficients $a_j$ and $b_j$ are given by
\begin{equation}
a_j=\omega_c\omega_j/D_j, ~b_j=\bigl(\omega_x^2-\omega^2_j\bigr)/D_j
\end{equation}
with
$D_j=\bigl[\bigl(\omega_x^2-\omega_j^2\bigr)^2+\omega_x^2\omega_c^2\bigr]^{1/2}$.
In the absence of SO coupling, the spectrum represents two ladders of
equidistant levels (for each spin projection) with energies
\begin{equation}
\label{eigen-energy}
E^{(0)}_{n_1n_2\pm}=\bigl(n_1+1\bigl/2\bigr)\omega_1
+\bigl(n_2+1\bigl/2\bigr)\omega_2\pm\omega_Z,
\end{equation}
where  $\omega_Z=\frac{1}{2}g^\ast\mu_BB$ is the Zeeman energy.

The SO term $\hat{H}_{SO}=\hat{H}_D+\hat{H}_R$ causes an admixture of
the oscillator states $|n_1n_2\rangle_\pm$ with different orbital ($n_j$)
and spin ($\pm$) quantum numbers. In a titled field, the calculation
of the SO matrix elements is carried out in two steps (see Appendix \ref{app:so}).
First, the spin operators $\bm{\sigma}$ in $\hat{H}_{SO}$ are rotated in
spin space to align the spin-quantization axis along the total field
{\bf B}. Second, the orbital operators $\bm{\pi}$ in $\hat{H}_{SO}$ are
expressed via new canonical variables $P_j$, $Q_j$. The expressions
for the general matrix elements are provided in Appendix \ref{app:so};
for the lowest adjacent levels with opposite spins
corresponding to the $\omega_2$ ladder, they have the form
%
%\begin{widetext}
\begin{eqnarray}
\label{MatrixElements1}
{}_\pm\langle 01\bigl|\hat{H}_{SO}\bigr|00\rangle_\mp
\qquad \qquad \qquad \qquad \qquad \qquad  \qquad \qquad
\nonumber\\
=
\frac{\alpha_R}{l_2\sqrt{2}}
\Bigl[\bigl(b_2\cos\theta\pm a_2\bigr)
\cos\varphi
+ i\bigl(a_2\cos\theta\pm b_2\bigr)\sin\varphi\Bigr]
\nonumber\\
-
\frac{i\alpha_D}{l_2\sqrt{2}}\Bigl[\bigl(a_2\cos\theta\mp b_2\bigr)
\cos\varphi
-i\bigl(b_2\cos\theta\mp a_2\bigr)\sin\varphi\Bigr],
\end{eqnarray}
%\end{widetext}
%
where $l_j=(m\omega_j)^{-1/2}$.
These matrix elements explicitly depend on the tilt, $\theta$, and
azimuthal, $\varphi$, angles as well as on the QD geometry encoded in
the coefficients $a_j,b_j$. In general, the magnitude of SO coupling
is small compared to the level separation, $\alpha\bigl/l_j\ll\omega_j$,
and, accordingly, the SO-induced level admixture is weak. However,
the level mixing gets strongly enhanced near the resonance, i.e., when
the spacing between adjacent levels is of the order of the SO energy:
$\omega_2-\omega_Z\sim \alpha_{R,D}/l_2$ (see Fig.~\ref{fig1}). This
can be achieved, e.g., by varying the Zeeman energy with the tilt
angle $\theta$. The corresponding anticrossing gap,
\begin{equation}
\Delta=2
\bigl|{}_+\langle 01\bigr|\hat{H}_{SO}\bigl|00\rangle_-\bigr|,
\end{equation}
is evaluated from Eq.\ (\ref{MatrixElements1}) as
%
%\begin{widetext}
\begin{eqnarray}
\label{gap1}
&&
\mbox{\hspace{-4mm}}
\Delta^2
\!
=
\!
\frac{\alpha_R^2}{l_2^2}
\Bigl[\bigl(b_2\cos\theta+a_2\bigr)^2
\!
\cos^2\varphi+\bigl(b_2+a_2\cos\theta\bigr)^2
\!
\sin^2\varphi\Bigr]
\nonumber\\
&&
\, \,
+
\frac{\alpha_D^2}{l_2^2}\Bigl[\bigl(b_2\cos\theta-a_2\bigr)^2
\!
\sin^2\varphi+\bigl(b_2-a_2\cos\theta\bigr)^2
\!
\cos^2\varphi\Bigr]
\nonumber\\
&&
\, \,
+
\frac{\alpha_R\alpha_D}{l_2^2}
\bigl(a_2^2+b_2^2\bigr)\sin^2\theta\sin 2\varphi.
\end{eqnarray}
%\end{widetext}
%
The gap magnitude is governed by the angle-dependent interference
between Rashba and Dresselhaus terms. Importantly, in elliptical QDs,
such interference depends on the dot {\em geometry} via the
coefficients $a_j$, $b_j$.

Thus, near the resonance, $\omega_2-\omega_Z\sim\Delta$, the energies
of the lowest excited states,
\begin{equation}
E_{2,3}=\omega_2\mp \frac{1}{2}\sqrt{(\omega_2-\omega_Z)^2+\Delta^2},
\end{equation}
acquire a strong angular dependence. At the same time, the
phonon-assisted transitions between these states $|j\rangle$ ($j=2,3$)
and the ground state $|1\rangle$ are enhanced due to the strong
admixture of constituent orbital levels (see Fig.~\ref{fig1}). As a
result, the spin relaxation becomes anisotropic with respect to the in-plane
field orientation $\varphi$. As we show below, the relaxation
exhibits anomalous sensitivity of to the system parameters in
a narrow range of $\varphi$ where the SO terms interfere destructively.

\section{Phonon-assisted spin relaxation}
\label{phonon}

The transition rate between state $|j\rangle$ and the ground state
$|1\rangle$ is given by
\begin{equation}
\Gamma_{1j}=2\pi\sum_{{\bf Q}\lambda} |
\langle 1 |U_\lambda |j\rangle |^2 \delta\bigl(E_1-E_j
+\omega_{{\bf Q}\lambda}\bigr),
\end{equation}
where the sum runs over acoustic phonon modes $\lambda$ with dispersion
$\omega_{{\bf Q}\lambda}=c_\lambda Q$, $c_\lambda$ being the sound
velocity, and 3D momenta ${\bf Q}=({\bf q},q_z)$. The transition
matrix element is a product of phonon and electron contributions,
\begin{equation}
\langle 1 |U_\lambda |j\rangle=M_\lambda ({\bf Q})
\langle 1 |e^{i{\bf Q}{\bf R}}|j\rangle,
\end{equation}
where the phonon part,
\begin{equation}
M_\lambda({\bf Q})=\Lambda_\lambda({\bf Q})+i\Xi_\lambda({\bf Q}),
\end{equation}
includes piezoelectric, $\Lambda_\lambda$, and deformation,
$\Xi_\lambda$, contributions.\cite{gantmakher-book} In the numerical
calculations below, we include both longitudinal and two transverse
piezoelectric acoustical modes.
The details can be found in Appendix \ref{app:phonon}.
The electron matrix element can, in turn, be decomposed
into a product of transverse  and in-plane contributions,
\begin{equation}
\langle 1 |e^{i{\bf Q}{\bf R}} |j\rangle=f_z (q_z )f_{1j} (\bf q).
\end{equation}
The transverse contribution $f_z (q_z )$ is determined by the 2D
confinement, assumed parabolic below, while the in-plane
contribution $f_{1j}(\bf q)$ is evaluated as follows.

Not too far from the resonance region,
$\bigl|\frac{\omega_2-\omega_Z}{\omega_2+\omega_Z}\bigr|\ll 1$,
it is sufficient to restrict calculations to the lowest four levels of the
$\omega_2$ ladder. The states $|2\rangle$ and $|3\rangle$,
corresponding to the anticrossing levels, are superpositions of
unperturbed states,
\begin{equation}
|j\rangle=d^{(j)}_{00-}|00\rangle_-+d^{(j)}_{01+}|01\rangle_+,
\end{equation}
while the ground state acquires a small admixture from the upper orbital
of opposite spin,
\begin{equation}
\bigl|1\bigr\rangle=d^{(1)}_{00+}\bigl|00\rangle_+
+d^{(1)}_{01-}\bigl|01\rangle_-.
\end{equation}
The coefficients $d$ are obtained as
\begin{eqnarray}
\label{coeff_d}
&&d^{(3)}_{01+}=d^{(2)}_{00-}=\bigl( 1+e^{-2\beta}\bigr)^{-1/2},
\nonumber\\
&&d^{(3)}_{00-}=- d^{(2)\ast}_{01+}=e^{i\eta}\bigl(1+e^{2\beta}\bigr)^{-1/2},
\nonumber\\
&&d^{(1)}_{00+}=\bigl(1+\xi^2\bigr)^{-1/2},
\,\,
d^{(1)}_{01-}=\xi^\ast \bigl(1+\xi^2\bigr)^{-1/2},
\end{eqnarray}
where
\begin{equation}
\xi=\frac{{}_-\bigl\langle
01\bigl|\hat{H}_{SO}\bigr|00\bigr\rangle_+} {\omega_2+\omega_Z}\sim
\frac{\alpha_{R,D}}{l_2(\omega_2+\omega_Z)}
\end{equation}
is the ratio of SO and orbital energies, while
$
\eta=\arg \bigl({}_+\langle 01\bigl|\hat{H}_{SO}\bigr|00\rangle_-\bigr)
$
is the phase of the SO matrix element. The parameter $\beta$ characterizes
the proximity to the resonance:
\begin{equation}
\sinh\beta=(\omega_2-\omega_Z)/\Delta
\end{equation}
is the detuning in units of the anticrossing gap.  In deriving
Eqs.~(\ref{coeff_d}), we neglected higher-order terms in the SO coupling
suppressed by the additional factor $\bigl|\xi\bigr|\ll 1$. In the
resonance region $\beta\sim 1$, the coefficients $d^{(2)}\sim
d^{(3)}\sim 1$, i.e., the levels 2 and 3 are well hybridized. Away
from the resonance, $\Delta \ll |\omega_2-\omega_Z|\ll
(\omega_2+\omega_Z)$, corresponding to $|\beta| \gg 1$, the
eigenstates 2 and 3 are mainly determined by the unperturbed states
$|00\rangle_-$ and $|01\rangle_+$ (and vice versa) to the left
(right) from the resonance region.

In terms of the coefficients $d$, the 2D matrix element takes the form
\begin{equation}
f_{1j}({\bf q})=d^{(1)}_{01-}d^{(j)\ast}_{00-} f_{0001}({\bf q})
+d^{(1)}_{00+} d^{(j)\ast}_{01+} f_{0100}({\bf q}),
\end{equation}
with
$
f_{m^\prime n^\prime mn}=\langle m^\prime n^\prime |e^{i{\bf qr}}
|mn\rangle.
$
The relevant elements are (see Appendix \ref{app:orbit})
\begin{equation}
f_{0001}=f_{0100}^\ast=\frac{il_2q}{\sqrt{2}}
e^{-(a_1^2l_1^2+a_2^2l_2^2)q^2/4}
(-a_2\cos\phi+ib_2\sin\phi),
\end{equation}
with $\phi=\arg (\bf q)$.
We then obtain
\begin{equation}\label{InPlane3}
f_{1j}=
-\frac{il_2q}{\sqrt{2}}
e^{-(a_1^2l_1^2+a_2^2l_2^2)q^2/4}
\bigl[F^{(j)}_1e^{-i\phi}+F^{(j)}_2e^{i\phi}\bigr],
\end{equation}
where
\begin{eqnarray}
\label{functions4_1}
F^{(j)}_1=\frac{1}{2}
\Bigl[
(a_2+b_2)d^{(j)}_{01+} d^{(1)\ast}_{00+}
+(a_2-b_2)d^{(j)}_{00-}d^{(1)^\ast}_{01-}
\Bigr],
\,\,
\nonumber\\
\label{functions4_2}
F^{(j)}_2=\frac{1}{2}\Bigl[
(a_2-b_2)d^{(j)}_{01+}d^{(1)^\ast}_{00+}
+(a_2+b_2)d^{(j)}_{00-}d^{(1)^\ast}_{01-}
\Bigr] .
\,\,
\end{eqnarray}
Substituting Eq.\ (\ref{InPlane3}) into the full transition matrix
element,
\begin{equation}
\langle 1 |U_\lambda |j\rangle=M_\lambda ({\bf Q}) f_z (q_z )f_{1j}
(\bf q),
\end{equation}
the scattering rate $\Gamma_{1j}$ is obtained by summing
up over phonon modes in a standard manner
\cite{khaetskii-prb00,lyanda-prb02,golovach-prl04,bulaev-prb05,ulloa-prb05}
(see Appendix \ref{app:phonon}).
The result is the product
\begin{equation}\label{scattering44}
\Gamma_{1j}=G_{1j}\bigl(\varphi\bigr)W_j, \ \ \ j=2,3
\end{equation}
where the geometric factor
\begin{equation}
G_{1j}=\bigl|F^{\left(j\right)}_1\bigr|^2+\bigl|F^{\left(j\right)}_2\bigr|^2
\end{equation}
is determined {\em only} by the SO-induced admixture of electronic states,
encoded in the coefficients $d$, while the phonon factor $W_j$
describes the probability of phonon-assisted transitions between levels
separated by energy
$E_j-E_1$ (see Appendix \ref{app:phonon}).
Note that $W_j$ are nearly independent of the SO coupling; in the
resonance region, the SO contribution to $W_j$ is $\sim \xi^2 \ll 1$
and can be neglected.

Thus, near the resonance, the dependence of scattering rate $\Gamma$
on the SO parameters and, accordingly, on the azimuthal angle
$\varphi$, comes only through the geometric factor
$G_{1j}$. Remarkably, this factor can be extracted directly from the
experimental data via the {\em differential} relaxation rate
normalized to its value at some angle (e.g., $\varphi=0$):
\begin{equation}\label{normalized1}
\frac{\Delta\Gamma_j}{\Gamma}=
\frac{\Gamma_{1j}\bigl(\varphi\bigr)-\Gamma_{1j}\bigl(0\bigr)}
{\Gamma_{1j}\bigl(0\bigr)}=
\frac{G_{1j}\bigl(\varphi\bigr)-G_{1j}\bigl(0\bigr)}
{G_{1j}\bigl(0\bigr)},
\end{equation}
where the r.h.s. is independent of the phonon contribution.

\section{Discussion and numerical results}
\label{results}

Below we present our numerical results for spin relaxation rates in a
GaAs QD. Calculations were performed for a parabolic confining potential
with $\omega_x=2$ meV and $\omega_y=1.33$ meV, corresponding to the
ellipticity $\omega_x\bigl/\omega_y=3\bigr/2$. For comparison, results
for the circular QD ($\omega_x=\omega_y=2$ meV) are presented too. We
choose the parabolic transverse confinement with $l_z=4$ nm. The
values of the SO constants, if not specified otherwise, in GaAs were taken
as $\alpha_R=5$ meV{\AA} and $\alpha_D=16.25$ meV\AA, while the phonon
parameters were taken from Ref.\ \onlinecite{gantmakher-book}
(see Appendix \ref{app:phonon}).

In Fig.\ \ref{fig1} we plot the lowest energy levels and relaxation
rates for both circular and elliptical QDs as a function of the in-plane
field $B_{||}$ at fixed $B_\perp =6.0$ T and $\varphi=0$.
For chosen parameters, the level anticrossings at $\omega_2=\omega_Z$,
indicated by circled regions, for elliptical QDs are achieved at lower
$B_{||}$ due to a weaker confinement along the $y$ axis [see
Fig.\ \ref{fig1}(a)]. The relaxation rates
$\Gamma_{12}$ and $\Gamma_{13}$  are plotted in Fig.\ \ref{fig1}(b). The
sharp increase in $\Gamma_{12}$ ($\Gamma_{13}$) is caused by a stronger
SO-induced admixture of the $\bigl|00\bigr\rangle_-$ and
$\bigl|01\bigr\rangle_+$ states as one approaches to the resonance
from the left (right). To the right of the resonance
($\omega_Z>\omega_2$), $\Gamma_{12}$ is dominated by the orbital
transition between states $\bigl|00\bigr\rangle_+$ and
$\bigl|01\bigr\rangle_+$; so is $\Gamma_{13}$ to the left of the
resonance ($\omega_Z<\omega_2$). The flat
$B_{||}$ dependence in these regions is because in a narrow 2D layer
the orbital wave functions depend only on $B_\perp$. Apart from the
magnitude of $\Gamma$, the overall behavior is similar for circular
and elliptical QDs.
\begin{figure}
\centering
\includegraphics[width=0.99\columnwidth]{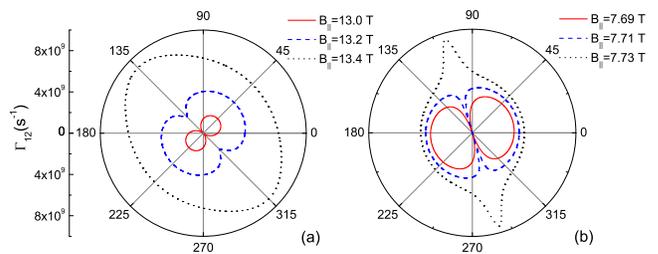}
\caption{\label{fig2}
(Color online) Angular dependence of the
relaxation rate $\Gamma_{12}$
for (a) circular and (b) elliptical QDs for $B_\perp =6.0$ T and several values
of $B_{||}$.
}
\end{figure}

In Fig.\ \ref{fig2}, we plot the relaxation rate $\Gamma_{12}$ as a
function of in-plane field orientation $\varphi$ at several values
of its magnitude $B_{||}$ in the resonance region. A strong 
azimuthal anisotropy is apparent: at certain angles, $\Gamma_{12}$
reaches minima that turn into maxima as $B_{||}$ sweeps through the
resonance. For a circular QD, the extrema of $\Gamma_{12}$ occur at
$\varphi=\pm\pi\bigl/4$ regardless of the values of $\alpha_R$ and
$\alpha_D$ [see Fig.\ \ref{fig2}(a)].  This anisotropy originates from
the angular dependence of the anticrossing gap
$\Delta\bigl(\varphi\bigr)$. Indeed, in the resonance region,
$|\omega_2-\omega_Z|\bigl/\Delta \sim 1$, the SO contribution to
$\Gamma_{12}$ takes the simple form
\begin{equation}
\label{G12}
G_{12}\bigl(\varphi\bigr)=
\bigl(a_2^2+b_2^2\bigr)
\Biggl(
1-\frac{\omega_2-\omega_Z}{\bigl[\Delta^2+\bigl(\omega_2-\omega_Z\bigr)^2\bigr]
^{1/2}}
\Biggr).
\end{equation}
For a circular QD, the expression (\ref{gap1}) for the gap simplifies to
\cite{falko-prb92,golovach-prl04}
\begin{equation}
\Delta^2=\frac{\omega_2}{\Omega}\bigl(\lambda_D^2+\lambda_R^2
+2\lambda_D\lambda_R\sin2\varphi\bigr)
\end{equation}
with
$
\lambda_D=\alpha_D\sin^2(\theta/2)/l_2
$
and
$
\lambda_R=\alpha_R\sin^2(\theta/2)/l_2,
$
so the extrema of $\Delta$ at
$\varphi=\pm\pi/4$ translate into the extrema of $G$. In contrast, in
elliptical QDs, the angular dependence of $\Gamma$ depends on the
system parameters [see Fig.~\ref{fig2}(b)]. The additional asymmetry
introduced by the QD ellipticity modifies the interference between
Rashba and Dresselhaus terms and shifts the extrema away from
$\pm\pi/4$. The extrema of $\Delta$, Eq.\ (\ref{gap1}), now depend on
both the QD ellipticity and the SO parameters, and occur at
\begin{equation}\label{condition1}
\tan 2\varphi_e=-2\frac{\alpha_R\alpha_D}{\alpha_R^2-\alpha_D^2}
\frac{a_2^2+b_2^2}{a_2^2-b_2^2}.
\end{equation}
For the parameters of Fig.~\ref{fig2}(b), $\varphi_e=106^{\rm o}$.
\begin{figure}
\centering
\includegraphics[width=0.8\columnwidth]{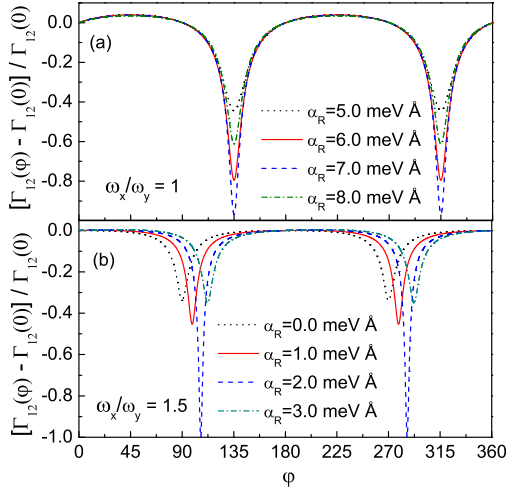}
\caption{\label{fig3}
(Color online)  Angular dependence of the differential
relaxation rates for (a) circular and (b) elliptical QDs for several
values of the coefficients $\alpha_R$ at $B_\perp=6.0$ T and
$B_{||}=13.2$ T (a) and $B_{||}=7.7$ T (b).
}
\end{figure}

Such interplay between QD geometry and SO interactions suggests
a way to {\em simultaneously} determine both SO couplings
and QD ellipticity from the measured angular dependence
of the differential relaxation rate $\Delta \Gamma/\Gamma$,
Eq.\ (\ref{normalized1}). In the resonance region, the phonon
contribution $W_j$ drops out, as does the prefactor in
Eq.\ (\ref{G12}), so $\Delta \Gamma/\Gamma$ is determined
solely by the anticrossing gap $\Delta(\varphi,\omega_x,\omega_y)$.
The angular dependence of $\Delta\Gamma\bigl/\Gamma$ is shown in
Fig.~\ref{fig3} for several values of Rashba coupling $\alpha_R$ which
can be varied, e.g., with an external electric field.\cite{winkler-book}
For a circular QD, change in $\alpha_R$ does not affect the minima positions
at $\varphi=-\pi\bigl/4$, as discussed above; however, the modulation
depth varies  strongly [see Fig.~\ref{fig3}(a)]. Note that the
$\alpha_R$ dependence is nonmonotonic; the deepest minimum occurs for
the almost complete  {\em destructive} interference of the SO terms at
$\alpha_R\bigl/\alpha_D\approx\tan^2(\theta/2)$.

Deviations from circular QD shape give rise to {\em angular
dispersion} of spin relaxation in the parameter space.  For elliptical
QDs, variations of $\alpha_R$ shift the minima
positions of $\Delta\Gamma\bigl/\Gamma$ [see Fig.~\ref{fig3}(b)].  In
fact, the sensitivity of $\Delta\Gamma\bigl/\Gamma$ to the system
parameters is drastically enhanced for $\varphi$ in the vicinity of the
critical angles $\varphi_e$, Eq.\ (\ref{condition1}), for which the
destructive interference between the Rashba and Dresselhaus terms is
strongest. Thus, a scan of the angular dependence of the experimental
differential relaxation rate in this narrow domain would enable an
unambiguous extraction of both SO and QD geometry parameters.

\section{conclusion}
\label{conc}
In summary, we proposed a method for simultaneous extraction of
both the spin-orbit constants as well as the quantum dot shape from the
angular anisotropy of the differential spin relaxation rate in a tilted
field. The underlying mechanism is based upon the enhanced
sensitivity of phonon-assisted spin-flip transitions to the system
parameters in the vicinity of level anticrossings. This sensitivity
arises from destructive interference between the SO terms in a narrow
domain of in-plane field orientations in the presence of asymmetric
confinement. Note that such interplay between SO interactions and QD
geometry cannot be captured by simplified descriptions of elliptical
QDs using circular QDs with effective parameters.\cite{tokura-nature02}

\section{acknowledgement}
This work was supported  by the NSF under Grant No. DMR-0606509 and by
the DoD under contract No. W912HZ-06-C-0057.

%\end{document}

\appendix
\section{Spin-orbit matrix elements in tilted field}
\label{app:so}
Here we describe calculation of the SO matrix elements in elliptical
QD in tilted magnetic field. It is convenient to work in the basis in
which Zeeman term is diagonal. Therefore, we choose the spin
quantization axis along the total field and perform the corresponding
rotation of the Pauli matrices:
\begin{eqnarray}
\label{transf-spin}
\sigma_x
&\rightarrow&
\sigma_x\cos\theta\cos\varphi-\sigma_y\sin\varphi+\sigma_z\sin\theta\cos\varphi,\nonumber\\
\sigma_y
&\rightarrow&
\sigma_x\cos\theta\sin\varphi+\sigma_y\cos\varphi+\sigma_z\sin\theta\sin\varphi,\nonumber\\
\sigma_z
&\rightarrow&
 -\sigma_x\sin\theta+\sigma_z\cos\theta .
\end{eqnarray}
The orbital variables are chosen to diagonalize the free Hamiltonian
$\hat{H}_0$ with the help of transformation
Eq.\ (\ref{transf-ellipt}). In this basis, the Hamiltonian 
$\hat{H}_0+\hat{H}_Z$ is diagonal in both orbital and spin spaces, 
with the eigenstates $|n_1,n_2\rangle_\pm$ corresponding to two
uncoupled oscillators with energies given by 
Eq.\ (\ref{eigen-energy}). The calculation of matrix elements of
$\hat{H}_{SO}$ between these eigenstates  is convenient to perform
by utilizing rising
$\hat{c}_j^\dagger=\frac{1}{\sqrt{2}}\Bigl(\frac{Q_j}{l_j}-il_jP_j\Bigr)$
and lowering
$\hat{c}_j=\frac{1}{\sqrt{2}}\Bigl(\frac{Q_j}{l_j}+il_jP_j\Bigr)$
operators. The corresponding non-zero matrix elements have the form
\begin{widetext}
\begin{eqnarray}
\label{MatrixElements2}
{}_\pm\langle n_1,n_2\bigl|\hat{H}_{SO}\bigr|n_1,n_2-1\rangle_\mp
=
\sqrt{\frac{n_2}{2}}\biggl(\frac{\alpha_R}{l_2}
\Bigl[\bigl(b_2\cos\theta\pm a_2\bigr)
\cos\varphi + i\bigl(a_2\cos\theta\pm b_2\bigr)\sin\varphi\Bigr]
\nonumber\\
+
i\frac{\alpha_D}{l_2}\Bigl[\bigl(-a_2\cos\theta\pm b_2\bigr)
\cos\varphi+i\bigl(b_2\cos\theta\mp a_2\bigr)\sin\varphi\Bigr]\biggr),
\nonumber\\
%%%%%
{}_\pm\langle n_1,n_2\bigl|\hat{H}_{SO}\bigr|n_1,n_2-1\rangle_\pm
=
\pm
\sqrt{\frac{n_2}{2}}
\biggl[\frac{\alpha_R}{l_2}
\bigl(b_2\cos\varphi+ia_2\sin\varphi\bigr)-i\frac{\alpha_D}{l_2}
\bigl(a_2\cos\varphi-ib_2\sin\varphi\bigr)
\biggr]\sin\theta
\nonumber\\
{}_\pm\langle n_1,n_2\bigl|\hat{H}_{SO}\bigr|n_1-1,n_2\rangle_\mp
=
\sqrt{\frac{n_1}{2}}\biggl(\frac{\alpha_R}{l_1}
\Bigl[\bigl(-b_1\cos\theta\mp a_1\bigr)
\cos\varphi \pm i\bigl(b_1\pm a_1\cos\theta\bigr)\sin\varphi\Bigr]
\nonumber\\
+
i\frac{\alpha_D}{l_1}\Bigl[\bigl(a_1\cos\theta\mp b_1\bigr)
\cos\varphi+i\bigl(-b_1\cos\theta\pm a_1\bigr)\sin\varphi\Bigr]\biggr),
\nonumber\\
%%%%%
{}_\pm\langle n_1,n_2\bigl|\hat{H}_{SO}\bigr|n_1-1,n_2\rangle_\pm
=
\mp
\sqrt{\frac{n_1}{2}}
\biggl[\frac{\alpha_R}{l_1}
\bigl(b_1\cos\varphi+ia_1\sin\varphi\bigr)-i\frac{\alpha_D}{l_1}
\bigl(a_1\cos\varphi-ib_1\sin\varphi\bigr)
\biggr]\sin\theta .
\end{eqnarray}
\end{widetext}
Note that $\hat{H}_{SO}$ does not couple states of different oscillators.
The second and fourth relations in Eqs.\ (\ref{MatrixElements2})
describe SO-induced transitions without spin-flip that are absent in
perpendicular field ($\theta=0$). Here, both Rashba and Dresselhaus
terms contribute to the coupling between same levels. In contrast,
for circular QD in perpendicular field with
$\omega_x=\omega_y=\omega_0$, the two SO terms only couple
different levels:
\begin{eqnarray}
{}_+\langle n_1,n_2\bigl|\hat{H}_{SO}\bigr|n_1,n_2-1\rangle_-
=\frac{\alpha_R}{l_2}\sqrt{\frac{n_2\omega_2}{\Omega}}\,
\nonumber\\
{}_-\langle n_1,n_2\bigl|\hat{H}_{SO}\bigr|n_1,n_2-1\rangle_+
=-i\frac{\alpha_D}{l_2}\sqrt{\frac{n_2\omega_2}{\Omega}},
\end{eqnarray}
where $\Omega=\sqrt{\omega_0^2+\omega_c^2/4}$ and
$\omega_{1,2}=\Omega\pm\omega_c/2$ (analogous relations hold for the
$\omega_1$ ladder).

\section{Electron-phonon matrix elements}
\label{app:phonon}
\subsection{Phonon matrix element $M(\bf Q)$}
The phonon-assisted transition rate is given by
\begin{equation}\label{Gamma_1j}
\Gamma_{1j}=2\pi\sum_{{\bf Q}\lambda}
| \langle 1 |U_\lambda |j\rangle |^2
\delta\bigl(E_1-E_j+c_\lambda Q\bigr)
\end{equation}
where the operator of electron-phonon interaction $U_\lambda$ has the
form \cite{gantmakher-book}
\begin{equation}\label{U_operator}
U_\lambda=\bigl(2\rho Qc_\lambda V\bigr)^{-1/2}
\bigl(eA_{{\bf Q}\lambda}+iQD_{{\bf Q}\lambda}\bigr)
e^{i{\bf Q}{\bf R}}\bigl(b_{{\bf Q}\lambda}+b^\dagger_{-{\bf Q}\lambda}\bigr).
\end{equation}
Here,
\begin{equation}
A_{{\bf Q}\lambda}=\frac{h_{14}}{Q^2\kappa}
\bigl(q_xq_ye^\lambda_z+q_xq_ze^\lambda_y+q_yq_ze^\lambda_x\bigr)
\end{equation}
is the amplitude of electric field created by phonon strain,
and deformation potential $D_{{\bf Q}\lambda}$ contains only
longitudinal acoustical (LA) component,
$D_{{\bf Q}\lambda}=\delta_{\lambda,LA}\Xi_0$, with
$\Xi_0$ being a constant of the deformation potential. Also,
$b^\dagger_{{\bf Q}\lambda}$ ($b_{{\bf Q}\lambda}$)
creates (annihilates) phonon  with dispersion
$\omega_{{\bf Q}\lambda}=Qc_\lambda$,
$V$ is the QD volume,  ${\bf R}=({\bf r},z)$ is 3D electron radius vector,
$\rho$ is the crystal mass density,
${\bf e}^\lambda$ is the phonon polarization vector,
$c_\lambda$ is sound velocity,
$h_{14}$ is bulk phonon constant,
$\kappa$ is the static dielectric constant.
Accordingly, the phonon part $M_\lambda({\bf Q})$ of the transition
matrix element $\langle 1|U_\lambda|j\rangle$
includes both piezoelectric $\Lambda_\lambda$ and deformation
$\Xi_\lambda$ contributions:
\begin{equation}\label{scattering2}
M_\lambda({\bf Q})=\Lambda_\lambda({\bf Q})+i\Xi_\lambda({\bf Q})
\end{equation}
with piezoelectric part containing LA
and two transverse acoustical (TA1, TA2) modes.
Since polarization directions are given as
\begin{eqnarray}\label{L1}
&&{\bf e}^L=\hat{x}\sin\theta_{\bf Q}\cos\phi+\hat{y}
\sin\theta_{\bf Q}\sin\phi+\hat{z}\cos\theta_{\bf Q},
\nonumber\\
&&{\bf e}^{TA1}=\hat{x}\cos\theta_{\bf Q}\cos\phi
+\hat{y}\cos\theta_{\bf Q}\sin\phi-\hat{z}\sin\theta_{\bf Q},
\nonumber\\\
&&{\bf e}^{TA2}=-\hat{x}\sin\phi+\hat{y}\cos\phi
\end{eqnarray}
[$\theta_{\bf Q}=\arcsin\bigl(q\bigl/Q\bigr)$,
$\phi=\arg(\bf q)$], one gets
\begin{equation}\label{scattering3}
\Xi_{LA}\bigl({\bf Q}\bigr)=\Xi_0A_{LA}\sqrt{Q},
\end{equation}
where, as mentioned above,
only $LA$ mode is present for $\Xi_\lambda$.
For $\Lambda_\lambda$ one obtains
\begin{eqnarray}\label{scattering4}
&&\Lambda_{LA}\bigl({\bf Q}\bigr)=\frac{3}{2}\Lambda_0A_{LA}
\frac{q^2q_z}{Q^{7/2}}\sin 2\phi ,
\nonumber\\
\label{scattering5}
&&\Lambda_{TA1}\bigl({\bf Q}\bigr)=
\frac{1}{2}\Lambda_0A_{TA}
\biggl(2\frac{q_z^2}{q^2}-1\biggr)\frac{q^3}{Q^{7/2}}
\sin 2\phi ,
\nonumber\\
\label{scattering6}
&&\Lambda_{TA2}\bigl({\bf Q}\bigr)=\Lambda_0A_{TA}
\frac{qq_z}{Q^{5/2}}\cos 2\phi .
\end{eqnarray}
In the above equations,
$\Lambda_0=eh_{14}/\kappa$, and
$A_\lambda=\bigl(2\rho Vc_\lambda\bigr)^{-1/2}$.
The sound velocities of the transverse and longitudinal modes
are $c_{TA1}=c_{TA2}\equiv c_{TA}$ and $c_{LA}$,
respectively. For GaAs, the parameter values are
$eh_{14}=0.14$ eV/\AA, $\Xi_0=7$ eV, $c_{LA}=5.14\times 10^3$ m/s,
$c_{TA}=3.03\times 10^3$ m/s, $\rho=5.31$ g/${\rm cm}^3$, and $\kappa =12.79$.

\subsection{Expressions for $W_j$}
To obtain Eq.\ (\ref{scattering44}), one has to perform integration
over phonon modes in Eq.\ (\ref{Gamma_1j}). The transition matrix element has the form
$\langle 1 |U_\lambda |j\rangle=M_\lambda ({\bf Q}) f_z (q_z )f_{1j}
(\bf q)$, where $M_\lambda ({\bf Q})$ is described in the previous
subsection, $f_z(q_z)$ is provided in the next section, and
$f_{1j}$ is given by Eq.\ (\ref{InPlane3}). Then, integration over
$q_z$ eliminates $\delta$-functions in Eq.\ (\ref{Gamma_1j}).
Subsequent $\phi$-integration eliminates terms linear in
$F^{\left(j\right)}_i$.
Remaining $q$-integration leads to Eq.\ (\ref{scattering44})
where the phonon factor $W_j$ is given by
\begin{eqnarray}\label{scattering66}
W_j=
\frac{1}{2\pi}\sum_{\lambda}
\int_0^{\frac{\Delta E_j}{c_\lambda}}
\frac{A_\lambda\bigl(q\bigr)q^3dq}{\sqrt{\Bigl(\frac{\Delta E_j}{c_\lambda}\Bigr)^2-q^2}}
\qquad \qquad \qquad
\nonumber\\
\times e^{-\bigl(a_1^2l_1^2+a_2^2l_2^2\bigr)q^2/2}
\Biggl|f_z\Biggl(\sqrt{\biggl(\frac{\Delta E_j}{c_\lambda}\biggr)^2-q^2}\Biggr)\Biggr|^2,
\end{eqnarray}
with summation running over all phonon modes: deformation ($Def$),
longitudinal and transverse acoustical:
\begin{eqnarray}\label{Deformation1}
&&A_{Def}\bigl(q\bigr)=\Xi_0^2 A_{LA}^2
\frac{2\bigl(\Delta E_j\bigr)^2}{c_{LA}^3},
\qquad\qquad
\nonumber\\
\label{LA1}
&&A_{LA}\bigl(q\bigr)=\frac{9}{4}\Lambda_0^2A_{LA}^2
\frac{c_{LA}^5}{\bigl(\Delta E_j\bigr)^6}
q^4\biggl[\biggl(\frac{\Delta E_j}{c_{LA}}\biggr)^2-q^2\biggr],
\qquad
\nonumber\\\
\label{TA1}
&&A_{TA}\bigl(q\bigr)=\Lambda_0^2A_{TA}^2
\frac{c_{TA}^4}{\bigl(\Delta E_j\bigr)^5}
q^2\biggl[
\biggl[\biggl(\frac{\Delta E_j}{c_{TA}}\biggr)^2-q^2\biggr]^{3/2}
\nonumber\\
&&\qquad \qquad
+\frac{1}{4}
\frac{c_{TA}}{\Delta E_j}
\biggl[2\biggl(\biggl[\frac{\Delta E_j}{c_{TA}}\biggr]^2-q^2\biggr)
-q^2\biggr]^2\biggr],
\qquad
\end{eqnarray}
where two transverse contributions are compacted as a single mode,
$\Delta E_j=E_j-E_1$, and zero temperature has been assumed.

\section{Electron matrix elements in elliptical QD}
\label{app:orbit} The form factor $f_z\bigl(q_z\bigr)$ is calculated
in the assumption that electron in the transverse direction is
frozen on the lowest level. Its explicit expression for the
transverse parabolic confinement
$V_z\bigl(z\bigr)=\frac{m}{2}\omega_z^2z^2$ is: $
f_z\bigl(q_z\bigr)=e^{-l_z^2q_z^2/4}$ with
$l_z=\bigl(m\omega_z\bigr)^{-1/2}$.

The explicit expression for the in-plane orbital matrix element,
$f_{m^\prime n^\prime mn}=\langle m^\prime n^\prime\bigr|e^{i{\bf
qr}} \bigl|mn\rangle$, is
\begin{eqnarray}
\label{MatrixElement2}
f_{m^\prime n^\prime mn}=
\sqrt{\frac{m!n!}{m^\prime !n^\prime !}}
e^{-(a_1^2l_1^2+a_2^2l_2^2)q^2/4}
\qquad \qquad \qquad \qquad
\nonumber\\
\times\biggl[\frac{il_1q}{\sqrt{2}}\bigl(a_1\cos\phi+ib_1\sin\phi\bigr)\biggr]^{m^\prime-m}
L_m^{m^\prime-m}\biggl(\frac{1}{2}a_1^2l_1^2q^2\biggr)
~~~
\nonumber\\
\times\biggl[-\frac{il_2q}{\sqrt{2}}\bigl(a_2\cos\phi+ib_2\sin\phi\bigr)\biggr]^{n^\prime-n}
L_n^{n^\prime-n}\biggl(\frac{1}{2}a_2^2l_2^2q^2\biggr),
~~
\end{eqnarray}
where $L_n^\alpha(x)$ is the associate Laguerre polynomial, and
$m^\prime \geq m$ and $n^\prime\geq n$. In deriving Eq.\
(\ref{MatrixElement2}), the following identity has been used
\begin{eqnarray}
\label{MatrixElement}
\langle m^\prime\bigr| e^{ilq[(a_1e^{-i\phi}
+a_2e^{i\phi})\hat{c}^\dagger+
%\nonumber\\
(a_2e^{-i\phi}+
a_1e^{i\phi})\hat{c}]}\bigl|m\rangle
\qquad \qquad
\nonumber\\
=\sqrt{\frac{m!}{m^\prime !}}
%\exp\Bigl[-\frac{\bigl(a_1+a_2\bigr)^2l^2q^2}{2}\Bigr]\nonumber\\
e^{-(a_1+a_2)^2l^2q^2/2}
\bigl[ilq\bigl(a_1e^{-i\phi}+a_2e^{i\phi}\bigr)\bigr]^{m^\prime-m}
\nonumber\\
\times L_m^{m^\prime-m}\bigl[(a_1+a_2)^2l^2q^2\bigr].
\qquad
\end{eqnarray}

%%%%%%%%%%%%%%%%%%%%%%%%%%%%%%%%%%%%%%%%%%%%%

\end{document}